**Approximate Negative-Binomial Confidence Intervals: Asbestos Fiber Counts**

DAVID BARTLEY[1]*, JAMES SLAVEN[2] and MARTIN HARPER[3]

[1]*Consultant, Physics and Statistics, 3904 Pocahontas Avenue, Cincinnati, Ohio, USA;*
[2]*Department of Biostatistics, Indiana University School of Medicine, Indianapolis, Indiana, USA;* [3]*Health Effects Laboratory Division, Exposure Assessment Branch, National Institute for Occupational Safety and Health, Morgantown, West Virginia, USA*

ABSTRACT

The negative-binomial distribution is adopted for analyzing asbestos-fiber counts so as to account for both the sampling errors in capturing only a finite number of fibers as well as the inevitable human variation in identifying and counting sampled fibers. A simple approximation to this distribution is developed for the derivation of quantiles and approximate confidence limits. The success of the approximation depends critically on the use of Stirling's expansion to sufficient order, on exact normalization of the approximating distribution, on reasonable perturbation of quantities from the normal distribution, and on accurately approximating sums by inverse-trapezoidal integration. Accuracy of the approximation developed is checked through simulation and also by comparison to traditional approximate confidence intervals in the specific case that the negative-binomial distribution approaches the Poisson distribution.

The resulting statistics are shown to relate directly to early research into the accuracy of asbestos sampling and analysis. Uncertainty in estimating mean asbestos-fiber concentrations given only a single count is derived. Decision limits (limits of detection ($LOD$)) and detection limits are considered for controlling false positive and negative detection assertions and are compared to traditional limits computed assuming normal distributions.

*Author to whom correspondence should be addressed.
**Tel: 1 513 652 4949; e-mail: bartleydavy@gmail.com**

INTRODUCTION

Airborne asbestos-fiber concentrations are often determined in terms of total fiber counts $n$ obtained manually by counting fibers on a microscope slide within a specific area $A$ (often 100 fields with total area 0.785 mm$^2$). The value $n$ differs from the true mean number $N$ of fibers per area $A$, since the number of fibers to be counted in the finite area $A$ varies around $N$. This variation follows the Poisson distribution. Furthermore, counter variation introduces uncertainty, leading to observations overdispersed relative to the Poisson distribution where the variance equals the mean. A great variety of count data have been analyzed in this situation using the negative binomial distribution as extension of the Poisson distribution. (For instance, see Cameron and Trevedi, 2013.) This is the approach adopted in this paper.

Specifically, given a singe count $n$, simple expressions for confidence limits enclosing the unknown $N$ at specified probability are derived. Such limits are vital in establishing exceedance or, conversely, an upper bound on the true mean count. Furthermore, confidence in the presence or absence of fibers can be quantified.

BACKGROUND

*Confidence intervals about N*

The earliest reported confidence intervals for asbestos fiber counting originated with an experiment at the Health and Safety Laboratory, Cricklewood, England (Ogden, 1982). The experiment provided data from 66 filters analyzed by about 10 counters with a variety of mean asbestos fiber counts $N$. The data were modeled with variance $\sigma^2$ approximated by:

$$\sigma^2 = N + s^2 N^2, \tag{1}$$

determining an estimate $s \approx 20\%$ representing the count variation at large $N$. The term in equation (1) linear in $N$ represents Poisson variation in the number of fibers that appear within the microscope fields searched.

This value for *s* is consistent with a recent evaluation from Harper *et al.*, 2013, who analyzed the capabilities of laboratory counters in the analysis of chrysotile fibers. Six slides were analyzed independently by 7 laboratories, using a single counter in each. The slides were relocatable, meaning that identical fields were examined by each counter and that Poisson sampling error was absent. One-way analysis of the data was carried out, assuming that (unknown) true counts for each slide were represented by consensus lab means and that inter-laboratory relative variances were approximately constant over the slides and therefore that the variance was estimable by their averages with 6*(7-1) degrees of freedom. The result was an inter-laboratory relative standard deviation equal to 22%. This figure is close to the intra-laboratory (in this case, inter-counter) estimate, 20%, found by Ogden, 1982. How these figures would vary with asbestos type is unknown.

Confidence limits were determined (Ogden, 1982) by identifying an approximate pivot (i.e., a function of *N* whose distribution is known and is independent of *N*). The function

$$pivot[n,N] \equiv (n-N)/\sqrt{N+s^2N^2} ,  \quad (2)$$

where *n* are measured counts, was found to only weakly depend on *N*. The distribution of the roughly 660 values of the pivot in the Cricklewood experiment was plotted as in Figure 1.

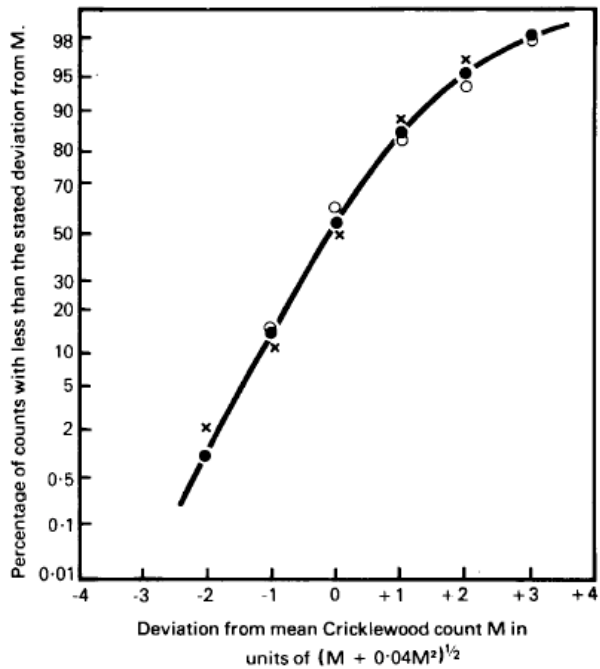

Fig. 1. Distribution of *pivot*[*N*] from the Cricklewood experiment (*M* = *N*). Solid circles include all the data, whereas open circles refer to N < 5, and "x "to N > 25 and indicate rough independence of the distribution on *N*. (UK Crown Copyright, 1982; reproduced from Ogden, 1982, by permission).

The graph is composed of a range of mean values *N*, and therefore both right and left single-sided 95%-confidence limits (−1.5, 2.0) can be read off from the graph. In other words, 95% of the measurements $\hat{n}$ have

$$(n-N)/\sqrt{N+s^2N^2} < 2.0 , \quad (3a)$$

and 95% have:

$$(n-N)/\sqrt{N+s^2N^2} > -1.5. \tag{3b}$$

Squaring equations (3) results in quadratic inequalities in *N*, which are easily solved for *N* in terms of functions, upper confidence limit *UCL*[*n*] and lower confidence limit *LCL*[*n*], of single-count values *n* as:

$$N < UCL[n] \text{ at } 95\% \text{ confidence}, \tag{4a}$$

and

$$N > LCL[n] \text{ at } 95\% \text{ confidence}, \tag{4b}$$

where

$$LCL[n] = \frac{2\hat{n} + 2.0^2 - \sqrt{(2n+2.0^2)^2 - 4(1-2.0^2 s^2)n^2}}{2(1-2.0^2 s^2)} \tag{5a}$$

$$UCL[n] = \frac{2n + 1.5^2 + \sqrt{(2n+1.5^2)^2 - 4(1-1.5^2 s^2)n^2}}{2(1-1.5^2 s^2)}. \tag{5b}$$

Note that NIOSH Method 7400 (NIOSH, 1994) and ASTM Method D7201 (ASTM, 2011) adopt the formalism presented above.

*Limit of detection (decision limit) and detection limit*

The limit of detection *LOD* is a figure defined to provide a means of limiting the false-positive rate (controlling "Type I errors", null-hypothesizing no significant asbestos fibers) in claiming the presence of a substance if actually absent or negligible. *LOD* can be defined as the bias-corrected signal level which is exceeded at probability < 0.1% when the substance is absent. Note that *LOD* as defined here is denoted a *decision limit* by Currie, 1984. Also, a *detection limit DL* may be defined as the mean asbestos fiber density which gives a signal > *LOD* at given confidence level (controlling false negatives (Type II errors)).

For many sampling/analytical methods the signal fluctuation approaches a constant as the presence of a substance -> 0. For example, in sampling unknown *M* (e.g., mass captured by a filter) the bias-corrected signal *m* may vary as:

$$m = M + \sqrt{\sigma_0^2 + M^2 s^2}\ \varepsilon, \tag{6}$$

where $\sigma_0$ is a constant, and where $\varepsilon$ varies about zero with a normal distribution and unit variance. With the variance of the signal $m$ given by:

$$Var[m] = \sigma_0^2 + M^2 s^2, \tag{7}$$

the constant $s$ is therefore the "true relative standard deviation" (i.e., the standard deviation relative to true values) for values of $M$ large enough that the fractional difference (equal to about $-\frac{1}{2}(\sigma_0/Ms)^2$) between $s$ and $\sqrt{\sigma_0^2 + M^2 s^2}/M$ is negligible. In the limit $M \to 0$, the constant $\sigma_0$ determines the fluctuation, and $LOD$ can be defined as:

$$LOD = 3 \times \sigma, \tag{8}$$

giving approximately 0.1% probability that $m > LOD$ if $M = 0$.

A value for $LOD$ was published by NIOSH, 1994, within NIOSH Method 7400 regarding asbestos fiber sampling and analysis using equation (8), even though equation (7) is not consistent with equation (1) at $N$ small enough that the fractional difference (equal to about $-100\% + sN^{1/2}$) between $Ns$ (from equation (7) at $\sigma_0 = 0$ and replacing $M$ by $N$) and $\sigma$ (equation (1)) is significantly negative. Furthermore, $LOD$ was defined as the signal from asbestos fibers *plus* background filter fibers needed to claim the presence of at least an asbestos fiber at high confidence. The value was therefore *not* bias-corrected.

Specifically, in the development of NIOSH 7400, many counts $n_i$ of interference fibers on blank filters indicated (with primes denoting densities (mm$^{-2}$)) that

$$\begin{aligned} Mean[n_i'] &= 2.5\ mm^{-2}\quad (N_i') \\ Variance^{1/2}[n_i'] &= 1.5\ mm^{-2}\ . \end{aligned} \quad \text{(Leidel, 1984))} \tag{9}$$

Then $LOD_{7400}$ was taken to equal:

$$\begin{aligned} LOD_{7400} &= N_i' + 3\ Variance^{1/2}[n_i'] \\ &= 2.5\ mm^{-2} + 3 \times 1.5\ mm^{-2} \\ &= 7\ mm^{-2}\ . \end{aligned} \tag{10}$$

The term with factor 3 is as in equation (8), while the value 2.5 $mm^{-2}$ refers to mean interference not related to the asbestos fibers sampled. Note that this value may be subtracted out from raw counts to give a bias-less signal if interfering fiber counts are stable. The quantitative difference between using equation (1) vs (7) is presented below in the section entitled APPLICATIONS.

PROPOSED

In this paper, we approach the issue of confidence intervals by adopting a specific distribution, namely the *negative binomial distribution* (equation (A1)), rather than an empirically determined distribution as measured by Ogden, 1982. As described in Appendix A, the negative binomial distribution is capable of describing the situation in which the variance $\sigma^2$ in a discrete variable $n$ is given by equation (1) and for which the Poisson distribution results if $s \to 0$ (Johnson et al., 1993). The negative binomial distribution then makes contact with the distribution of Ogden, 1982, and is adopted here for analyzing the distribution of counts. Johnson *et al.*, 1993, and Hilbe, 2011, both discuss the statistical adoption of the negative binomial distribution for modeling discrete random variables that ideally reflect the Poisson distribution, but which exhibit over-dispersion.

In order to obtain useful expressions for the confidence intervals, an approximation to the negative binomial distribution is worked out. The details of this derivation are generally not of interest to those applying the results and are therefore relegated to Appendix A. The body of the paper then summarizes results and the accuracy of the confidence intervals derived.

CONFIDENCE INTERVALS

This paper develops simple single-sample confidence intervals surrounding population-mean counts small enough that vestiges of the Poisson distribution are present. If larger counts are encountered, counting may be stopped at fewer than 100 fields. For example, Roggli *et al.*, 1992, suggests stopping at 200 fibers. In such cases, Poisson-distribution effects are negligible, and the relative standard deviation is approximately constant. This situation is not discussed further in the paper.

*Approximate quantiles of the negative binomial distribution*

We first consider simply the distribution quantile, namely the upper limit on a single count $n$ at specified probability $\beta$ (e.g., 95%) given measured counter variability $s$ and specified mean count *N*. This limit may be defined for discrete (rather than continuous) random variables as follows. The (upper) quantile $n_\beta[N,s]$ is the smallest integer for which

$$\Pr[n \leq n_\beta] \geq \beta. \tag{11}$$

As derived in Appendix A, the quantile $n_\beta$ for the negative binomial distribution after smoothing discrete discontinuities may be approximated simply as (equation (A22)):

$$(n_\beta - N)/\sigma \approx z_\beta + \tfrac{1}{6}(z_\beta^2 - 1)(1 + 2Ns^2)/\sigma, \qquad (12)$$

where $\sigma$ is given by equation (1), and where $z_\beta$ is the normal-distribution quantile (e.g., $z_\beta = -1.645$ at $\beta = 0.05$ and $z_\beta = +1.645$ at $\beta = 0.95$).

*Comparison of exact and approximate quantiles*

So as to illustrate accuracy, the above approximating functions were plotted along with exact functions, computed numerically, in a number of circumstances. Deviation of the quantile from mean count $N$ were plotted vs $N$ and normalized as by Ogden, 1982 and as in equation (2) by dividing by $\sqrt{N + s^2 N^2}$ so as to plot a reasonable range on a single figure.

Figure 2 shows 5% and 95% quantiles at *true relative standard deviation s* equal to 20% (as with data of Ogden, 1982). Figure 3 shows quantiles at $s$ = 40%. Figure 4 presents comparisons for the Poisson distribution ($s$ = 0).

Naturally, the approximations improve with decreasing $s$, considering the reliance on expansions (equation (A16)) about the mode of the probability distribution. Regardless, the accuracy is surprising in view of the simplicity of the approximating functions. Note that for some applications, only the averaging (indiscrete) quantiles of equation (A22) (also indicated in the figures) may be needed.

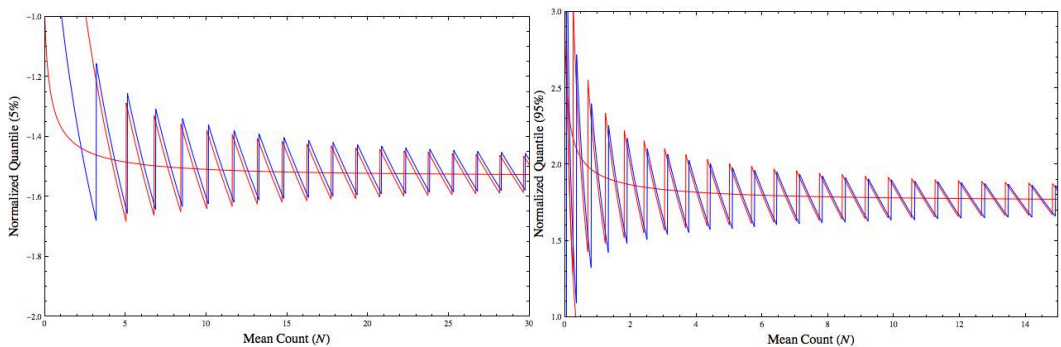

Fig. 2. Normalized negative-binomial quantiles---$(n_\beta - N)/\sqrt{N + s^2 N^2}$ ---for large-$N$ true relative standard deviation $s$ = 20% at levels β = 5% and 95% plotted vs population mean count $N$. Exact values (blue (color available in the online version of this paper)); approximations (red, equation (A21)) shifted slightly left for

graphical visibility; smooth curve (red, equation (A22)) neglecting the
discontinuities of the discrete values.

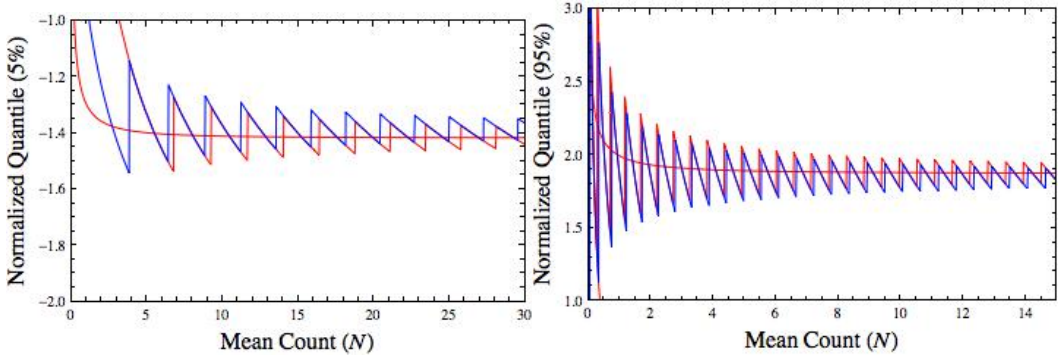

Fig. 3. Normalized negative-binomial quantiles at $s$ = 40% at levels $\beta$ = 5% and 95%
plotted vs population mean count $N$.  Exact values (blue); approximations (red)

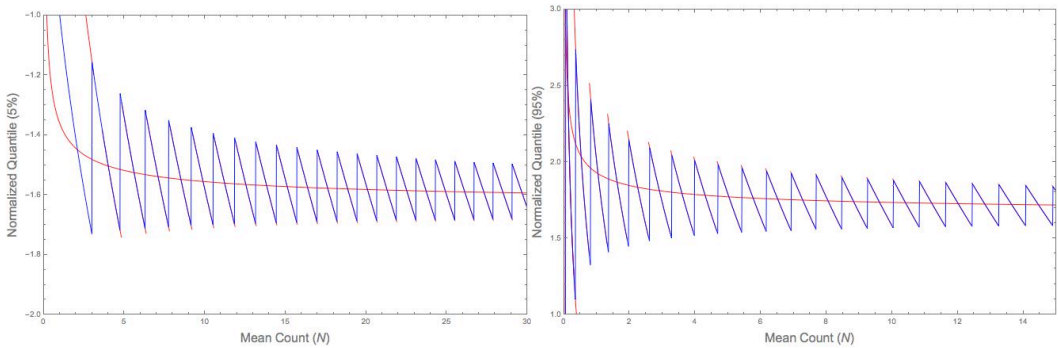

Fig. 4. Normalized Poisson distribution quantiles at levels $\beta$ = 5% and 95% plotted
vs population mean count $N$.  Exact values (blue); approximations (red).

*Single-sided confidence limits*

Given the above simple expressions for the distribution quantile, we now consider
confidence intervals limiting the true but unknown mean count $N$.  Because of the
discrete character of counts $n$, (just as with the Poisson distribution) exact
confidence limits at specified confidence do not exist.  However, approximate limits
result from the smoothed expression of equation (12), similar to the chi-square
approximation for the Poisson distribution (Johnson et al., 1993).  Equation (12)
indicates that $n_\beta$ is an increasing function of $N$.  Therefore, a single-sided confidence
limit on $N$ can be determined by solving for $N[n_\beta, s]$.  Then equation (11) implies
that

$$\Pr[N > N[n,s]] \approx \beta. \tag{13}$$

Note that the count $n$, not $N$, is the random variable in equation (13), expressing the sense of a *lower confidence limit* $LCL[n]$:

$$LCL[n] = N[n,s]. \tag{14}$$

The function $N[n_\beta, s]$ can be easily obtained by solving equation (12) for $\sigma^2$, resulting in a quadratic equation in $N$ upon using equation (1).

Yet simpler and useful is to break up the domain of interest into cases $s = 0$ (corresponding to the Poisson distribution) and $s > 0$ where an approximate pivot may be adopted (a function of $N$ and $n$ with distribution known and independent of $N$). Ogden, 1982, considered as approximate pivot the function $(n - N)/\sigma[N]$ (equation (2)), with $\sigma^2$ given by equation (1) and where then ideally, with quantile $n_\beta$, $(n_\beta - N)/\sigma$ would be nearly independent of $N$.

The constancy of this function within the above approximation to the negative binomial distribution was determined by using the smoothed quantile $\bar{n}_\beta$ (equation (A22)) and plotting $(\bar{n}_\beta - N)/\sigma$ vs $N$ at various values of $s$ (namely $s$ = 0%, 20%, and 40%). The results are shown in Figure 5.

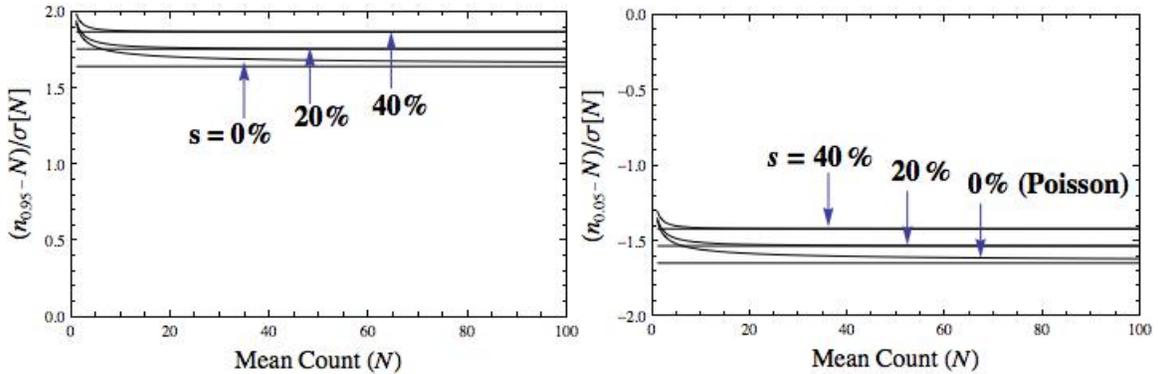

Fig. 5. Normalized 95%-quantiles vs the mean $N$ at several values of true relative standard deviation $s$. The horizontal lines represent asymptotic values $l_\beta$ at $N \to \infty$.

Note that as $N$ becomes large, equation (12) implies that

$$(n_\beta - N)/\sigma \to z_\beta + \tfrac{1}{3}(z_\beta^2 - 1)s \equiv l_\beta, \quad N \gg 1, \tag{15}$$

which clarifies the approximate pivot of equation (2) (Ogden, 1983). Figure 5 indicates that the normalized quantile (the left-hand side of equation 15)) rapidly approaches the limit $l_\beta$ as $N$ increases if $s \geq 20\%$. This means that $N[n_\beta, s]$ may be approximated by simply solving equation (15) for $N$ (by squaring both sides and solving the quadratic equation). The result, eliminating a spurious root, is:

$$LCL[n] = \frac{2n + l_\beta^2 - \sqrt{(2n + l_\beta^2)^2 - 4(1 - l_\beta^2 s^2)n^2}}{2(1 - l_\beta^2 s^2)}. \qquad (16a)$$

The same reasoning yields an upper single-sided confidence limit $UCL[n]$ at probability, $\beta$ (e.g., 95%):

$$UCL[n] = \frac{2n + l_{1-\beta}^2 + \sqrt{(2n + l_{1-\beta}^2)^2 - 4(1 - l_{1-\beta}^2 s^2)n^2}}{2(1 - l_{1-\beta}^2 s^2)}, \qquad (16b)$$

noting the sign change in front of the square root. The above expressions also result in 2-sided confidence intervals in the sense that:

$$\Pr[LCL[n] < N < UCL[n]] \approx 2\beta - 1 \text{ (e.g., 95\%, if } \beta = 0.975). \qquad (17)$$

The Poisson distribution (at $s = 0$) is obtained using equation (12) directly. Details are given in Appendix A. Comparison is made to traditional confidence intervals involving a relation to the chi-square distribution.

SIMULATIONS

In order to judge the accuracy of the above expressions when applied to discrete random variables, many simulations were carried out. At each of mean number $N$ = 5.1, 5.1, ..., 30, a random number generator produced 10,000 negative-binomial-distributed values for checking the above inequalities. Results are given in Figures (6,7) for $s = 20\%$ and $s = 40\%$.

If strict confidence limits existed, the data would fall close to horizontal straight lines. The scatter about these lines is evident in Figures (6,7). However, the scatter is strictly limited. In fact, the limits $\lim[N]$ in the single-sided case may be approximated by considering how the confidence level $\beta$ must change in order to increase the quantile $n_\beta$ of equation (12) by 1. The result is that:

$$\lim[N] \approx \beta \pm \frac{1/2}{(1 + \frac{1}{3} z_\beta s) \sqrt{2\pi} \, Exp[z_\beta^2 / 2] \, \sigma[N]}, \qquad (18)$$

where $\sigma[N]$ is given by equation (1). The 2-sided case of equation (18) is approximated by summation, as the upper and lower frequencies are generally not commensurate.

Note that though scatter is significant, no bias is evident in the confidence levels. The effect of discrete scatter on confidence limits can be estimated by computing the shifts in equations (16) induced by shifts in $\beta$. For example, at *s* = 20% and *N* = 10, the upper confidence limit shifts by only about $\pm 5\%$.

A further set of simulations was carried out, solving equation (12) for *N* exactly, rather than using the pivot as in equation (15). The results differ insignificantly from Figures (6,7). We conclude that at least for $s \geq 20\%$, the simpler expressions of equation (16) are adequate.

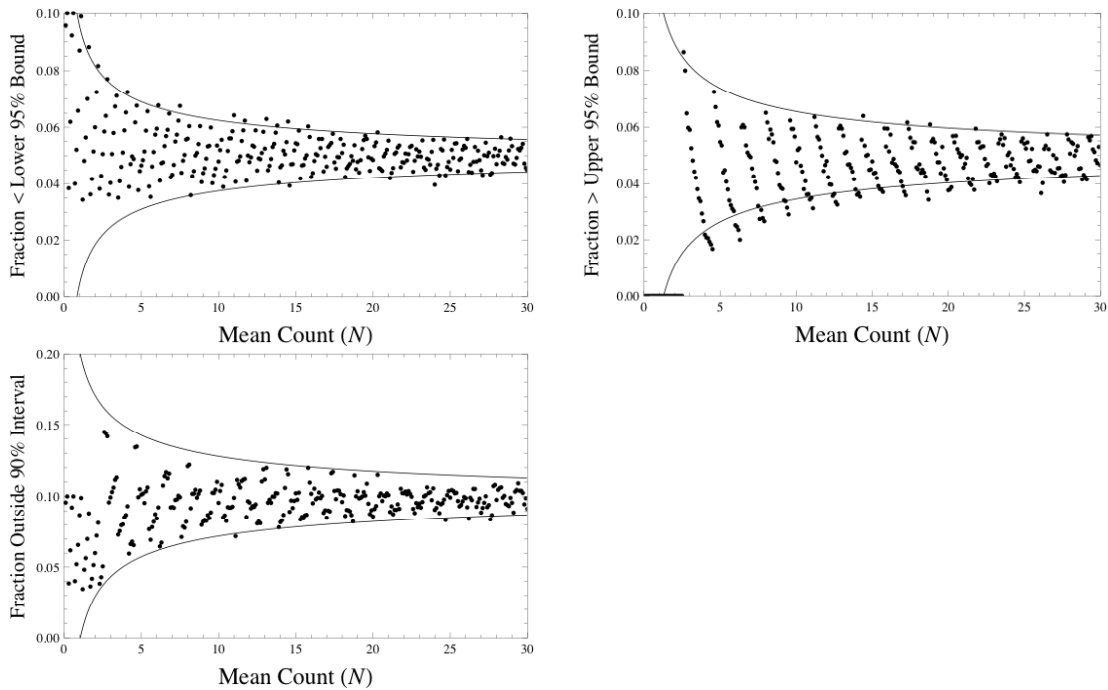

Fig. 6. Negative binomial confidence interval levels at *s* = 20% at nominal single-sided levels = 5% and 2-sided interval with nominal level = 10%. Solid curves are approximate limits on the discrete variation (equation (18)).

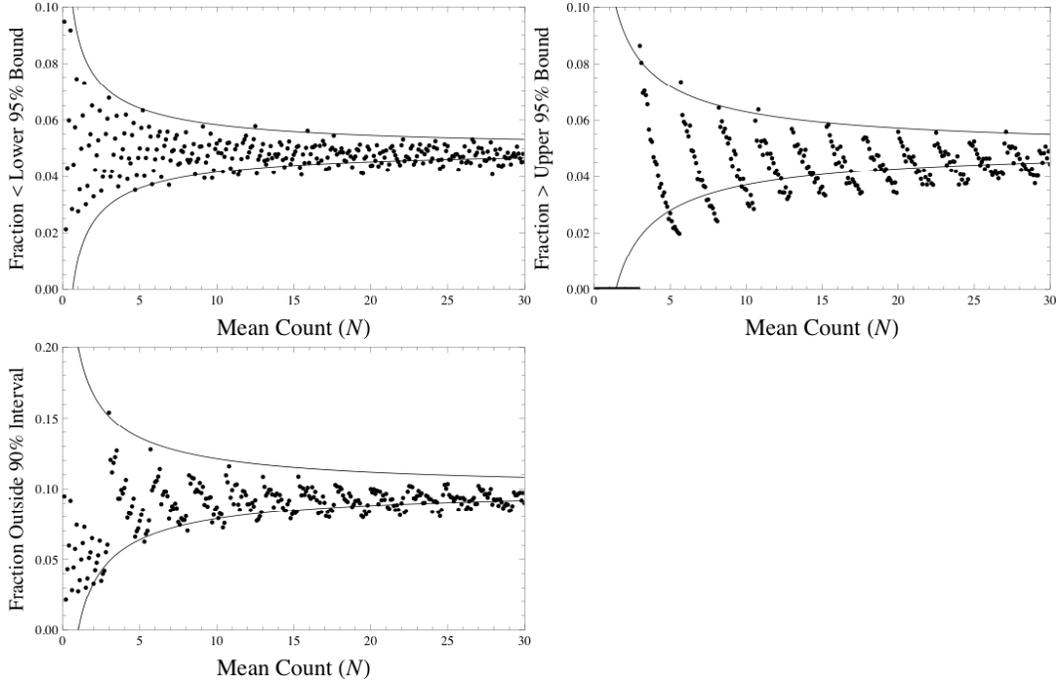

Fig. 7. Negative binomial confidence interval levels at *s* = 40%. Solid curves are approximate limits on the discrete variation (equation (18)).

APPLICATIONS

*Derivation of the early intuitive confidence limits*

One application of the negative-binomial formulism is derivation of the results published by Ogden, 1982. This is done directly using equation (15). At *s* = 20%, equation (15) indicates that

$$(n_{0.95} - N)/\sigma \to z_{0.95} + \tfrac{1}{3}(z_{0.95}^2 - 1)s = 1.8 \text{ , and} \qquad (19)$$

$$(n_{0.05} - N)/\sigma \to z_{0.05} + \tfrac{1}{3}(z_{0.05}^2 - 1)s = -1.5 \text{ .}$$

The limits (−1.5, 1.8) are consistent with the published values (−1.5, 2.0), noting that very few experimental data points were available at large mean count *N*. Agreement with experiment thus lends support to the utility of the negative-binomial distribution.

*Uncertainty*

Limits (–1.8, 2.1) at the 97.5% level may be obtained by replacing 1.645 by 1.960 in equation (19)).  This results in 95%-confidence limits given by equations (16) as indicated in Figure 8.

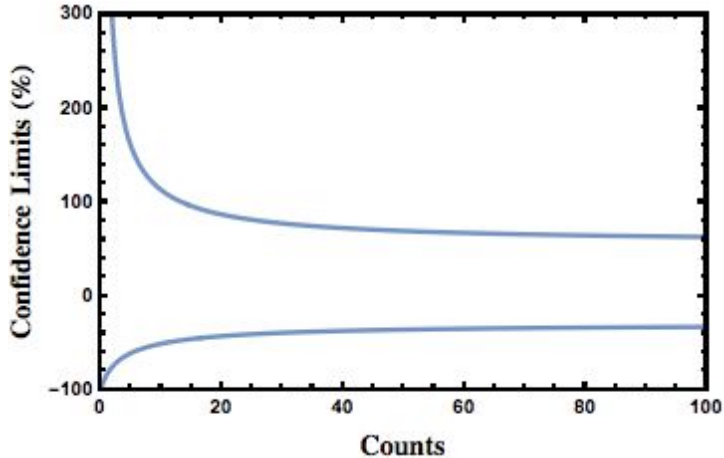

Fig. 8. Relative (i.e., $(n_\beta - n)/n \times 100\%$) confidence limits at the 95% level at intra-lab imprecision $s$ = 0.20.

The expanded uncertainty (ISO, 1995) in an estimate of $N$ on the basis of a single count can be read directly from this figure, with the understanding that the uncertainty is unusual in its asymmetry about zero.  Note that with interfering fibers present on sampling filters, fibers present and variable on filters (e.g. blanks) prior to sampling which may affect false positive rates, the above confidence limits refer to *total* mean fiber numbers in a given area, rather than specifically asbestos fibers. A table of lower and upper 2-sided 95%-confidence limits is given in Table 1 (and the values can be compared with the similar table in HSG248, 2005).

Table 1. Lower and upper 2-sided 95%-confidence limits (CL). Standard deviation (RSD) relative to count approaches 20% in large-count limit.  Values at count < 5 as derived from an asymptotic continuum model are only suggestive.

| Count | RSD (%) | Lower CL | Upper CL |
| --- | --- | --- | --- |
| 1 | 102 | 0 | 6 |
| 3 | 61 | 1 | 10 |
| 5 | 49 | 2 | 13 |
| 7 | 43 | 3 | 16 |
| 10 | 37 | 5 | 21 |
| 20 | 30 | 11 | 37 |
| 50 | 24 | 32 | 85 |
| 100 | 22 | 67 | 163 |
| 200 | 21 | 137 | 319 |

*Limit of detection (decision limit) and detection limit*

Specification of *LOD* can be improved over NIOSH Method 7400 by using the negative binomial distribution rather than the normal approximation of equations (8,10). Description of the experimental design that resulted in the variance in equation (9) no longer exists. For instance, it is not known whether exactly 100 fields were always examined or whether the filters came from different batches. The mean value, however, given in equation (9) is less problematic. In the following, the above mean value is accepted, whereas the variance is computed assuming intra-laboratory relative standard deviation for large counts is approximated as 20% as from the experiments of Ogden, 1982.

The standard deviation arising from considering count variance in a sample of area $A$ may be calculated as:

$$Variance[n_a + n_i] = (N_a + N_i) + s^2(N_a + N_i)^2, \qquad (20)$$

with $s$ = 20%, and with the mean background count density $N_i'$ taken as 2.5 $mm^{-2}$, as in Equation (20). Setting $N_a$ to zero and $A$ = 0.785 $mm^2$ for 100 fields then results in

$$\begin{aligned}Variance^{1/2}[n_i'] &= \sqrt{N_i'/A + s^2 N_i'^2} \\ &= 1.85 \ mm^{-2}\end{aligned}, \qquad (21)$$

a result not inconsistent with the value 1.5 $mm^{-2}$, especially if a variety of filter areas was sometimes assessed. The variation, filter to filter, in the number of interfering fibers was evidently found insignificant relative to the uncertainty in reading them. Henceforth, the figure 1.5 $mm^{-2}$ is ignored aside from indicating consistency.

The cumulative negative-binomial distribution and normal distribution giving Equation (20) are illustrated in Figure 9 at $N_a' = 0$, *TRSD* = 20%, $N_i'$ = 2.5 $mm^{-2}$, and $A$ = 0.785 $mm^2$.

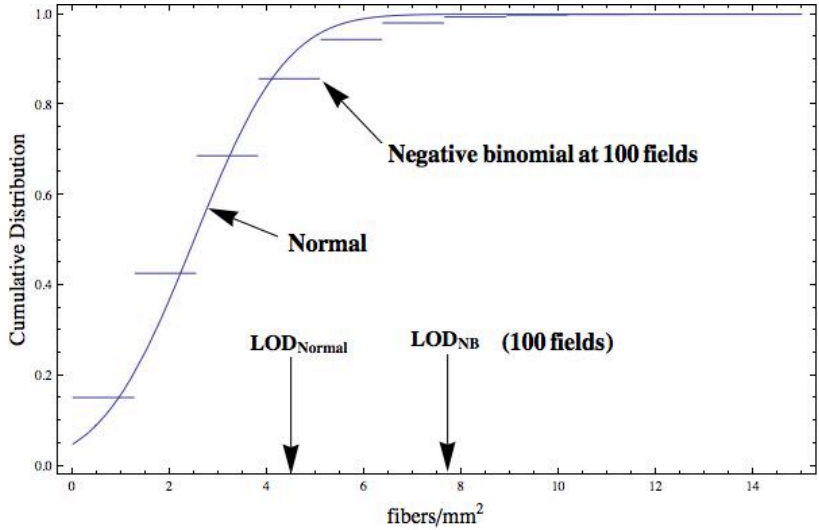

Fig. 9. Cumulative normal and negative-binomial distributions of background counts in 100 fields, assuming the mean value $N'_i = 2.5\ mm^{-2}$ as published in NIOSH 7400 and intra-counter variability = 20%. The *LOD* values indicated refer to bias-less signals from a single filter, i.e., with $2.5\ mm^{-2}$ subtracted out.

Now the quantile at confidence equal to 0.999 is easily computed for the negative-binomial distribution with the result:

$$(n'_i A)_{0.999} = 8. \tag{22}$$

Thus, finally, *LOD* is given by:

$$\begin{aligned} LOD &= (n'_i A)_{0.999} / A - N'_i \\ &= 7.7\ mm^{-2} \quad (at\ A = 0.785\ mm^2)\ . \end{aligned} \tag{23}$$

This value compares to

$$LOD_{normal} = 4.5\ mm^{-2} \tag{24}$$

(which is identical to $LOD_{7400} = 7\ mm^{-2}$, remembering that the values in equations (24,25) refer to bias-less signals (unlike the published NIOSH value)). The meaning of *LOD* is that the probability < 0.001 that the signal > $7.7\ mm^{-2}$ in the case of negligibly sampled asbestos. In other words, the false-positive rate in claiming the presence of asbestos is < 0.001 if $7.7\ mm^{-2}$ is taken as the threshold value for making such a claim. Note that the claimed confidence level equal to 0.999 may be misleading in view of uncertainty in the distribution of interfering non-asbestos fibers present on the sampling filters.  How to handle this source of uncertainty in terms of prediction intervals is described in ISO 15767 (2009).

Analysis using the negative-binomial distribution therefore is seen to give a qualitatively distinct result from the usual normal assumptions for computing the limit of detection.  Calculation of a detection limit *DL* (for controlling false negatives) would similarly differ.  Calculation as above indicates that *DL* = 12 mm$^{-2}$ (about 10 counts in 0.785 mm$^2$) at 80% confidence.

CONCLUSIONS

A simple approximation to the negative-binomial distribution has been developed.  The aim has been the statistical analysis of asbestos-fiber counts accounting for Poisson variation associated with errors in sampling a finite number of fibers as well as counter variability in analysis.  This approach has been shown to connect with early studies of count statistics reported by Ogden, 1982.  Estimation of the uncertainty in measuring a mean fiber concentration given a single count has been established.  Specification of upper or lower confidence limits on fiber concentrations is also possible, relevant to proving that an occupational exposure limit has or has not been exceeded.

Decision and detection limits for specifying false positive and negative detection rates are also possible.  Traditional specification using the normal distribution is shown to differ significantly from accounting for the discrete sampling problem.  For example, the decision limit computed from the negative-binomial distribution is found to be almost twice as large as from the normal distribution.

The negative-binomial approximation is so simple that other applications are no doubt possible.  For example, perhaps analysis of how best to use field blanks so as to minimize the effect of spurious fibers is feasible.  The variety of possible applications and corresponding experimental designs is large, as seen in disparate approaches of sampling and analysis described in NIOSH 7400, ASTM 7201, HSG 248, and ISO 8672.  Simplicity in the statistical analysis is vital.

**FUNDING**

Health Effects Laboratory Division, National Institute for Occupational Safety and Health, Centers for Disease Control and Prevention (Project No. 927001Q) to D.B. and J.S.

# Appendix A. NEGATIVE-BINOMIAL APPROXIMATION

*Parameterization for counting particles*

A simple and yet accurate closed-form approximation to quantiles of the negative-binomial distribution for an important range of parameters is derived here. The probability distribution function $P[n;p,r]$ for random integer $n$ distributed according to the negative-binomial distribution is given by:

$$P[n;p,r] = \frac{(n+r-1)!}{(r-1)!\,n!}(1-p)^r p^n. \tag{A1}$$

The parameters $r$ and $p$ may be any real numbers with $0 < p < 1$ and $r > 0$.

The mean and variance are given by:

$$\begin{aligned} Mean[n] &= pr/(1-p) \\ Variance[n] &= pr/(1-p)^2 \end{aligned} \tag{A2}$$

An alternative set of useful parameters $N$ and $s$ may be defined by:

$$\begin{aligned} p &= (Ns/\sigma)^2 \\ r &= s^{-2}, \end{aligned} \tag{A3}$$

where $\sigma^2$ is the variance:

$$\sigma^2 \equiv Variance[n] = N + s^2 N^2, \text{ and} \tag{A4a}$$

$$Mean[n] = N. \tag{A4b}$$

The parameter $s$ therefore represents a *true relative standard deviation* (relative to the mean $N$) in the limit $N \gg 1$. Also, the approach to the Poisson distribution in the limit $s \to 0$ ($r \to \infty$) is manifest. Application is therefore suggested, for example, in describing counts $n$ of mean $N$ particles per given area, with counter variability expressed by $s$.

The range, $s < 100\%$, is of practical importance. For example, Ogden, 1982, presented data on intra-lab asbestos counter variability with $s$ of the order of 20%. Similarly, NIOSH 7400 assumes a value $s \sim 45\%$ for covering inter-lab variability.

Therefore, analysis of the negative-binomial distribution with $r \gg 1$ is of interest. Also, the range with counts $n \gg 1$ is important. This suggests treating all the factorial arguments in equation (A1) asymptotically. By focusing on these ranges, upper single-sided quantiles may be accurately approximated for $N$ as small as 1.00, and, surprisingly, lower quantiles for $N > 3$ at $s < 60\%$.

*Asymptotic representation of the distribution function*

The approach taken here is to use Stirling's approximation for the factorials in equation (A1):

$$Log[z!] \approx z\,Log[z] - z + \tfrac{1}{2} Log[2\pi z] \quad for\ z \gg 1\,. \tag{A5}$$

Sums over discrete values of $n$ for defining the quantiles are approximated by integrals, adjusting end-points trapezoidally. Finally, perturbation from quantiles of the normal distribution is easily calculated.

Therefore, equation (A1) is approximated as:

$$\begin{aligned}
Log[P[n;p,r]] = &(n+r-1)Log[n+r-1] - (n+r-1) + \tfrac{1}{2} Log[2\pi(n+r-1)] \\
&-n\,Log[n] \quad\quad +n \quad\quad -\tfrac{1}{2} Log[2\pi n] \\
&-Log[(r-1)!] \\
&+r\,Log[1-p] + n\,Log[p]\,.
\end{aligned} \tag{A6}$$

The expression $Log[P[n;p,r]]$ is now expanded as a Taylor's series about its maximum, the peak of $P[n;p,r]$ itself. The first derivative is

$$\frac{d}{dn} Log[P[n;p,r]] = Log[n+r-1] + \tfrac{1}{2}(n+r-1)^{-1} - Log[n] - \tfrac{1}{2} n^{-1} + Log[p]\,. \tag{A7}$$

Treating $n$ as continuous, $P[n;p,r]$ therefore has a maximum at $n \approx n_0$ given, after collecting terms and exponentiation, by:

$$p\bullet(n_0 + r - 1)/n_0 = Exp[\tfrac{1}{2} n_0^{-1} - \tfrac{1}{2}(n_0 + r - 1)^{-1}]\,. \tag{A8}$$

At fixed r, as $n_0 \to \infty$ the argument of the exponential function $\to 0$ rapidly, proportional to $n_0^{-2}$, and the right-hand-side of Equation (A8) correspondingly $\to 1$. Therefore in an asymptotic limit, to lowest order, $n_0$ can be approximated by $n_{00}$ given by:

$$p\bullet(n_{00} + r - 1)/n_{00}] = 1\,, \tag{A9}$$

and therefore

$$n_{00} = p(r-1)/(1-p)$$
$$= N(1-s^2) \ . \tag{A10}$$

Iteration of equation (A8) then gives $n_0$ approximately according to:

$$p(n_0 + r - 1)/n_0] = 1 + \tfrac{1}{2}n_{00}^{-1} - \tfrac{1}{2}(n_{00} + r - 1)^{-1}$$
$$= 1 + \tfrac{1}{2}(1-p)/n_{00} \ . \tag{A11}$$

Thus, $n_0$ is approximated as simply:

$$n_0 \approx n_{00} - \tfrac{1}{2}$$
$$= N(1-s^2) - \tfrac{1}{2}. \tag{A12}$$

With a view to expanding $Log[P[n;p,r]]$ as a Taylor's series about $n_0$, the 2nd and 3rd derivatives are:

$$\frac{d^2}{dn^2} Log[P[n;p,r]] = (n+r-1)^{-1} - n^{-1} - \tfrac{1}{2}(n+r-1)^{-2} + \tfrac{1}{2}n^{-2}$$
$$\approx -(1-p)/n_{00} \quad at \ n = n_0 \tag{A13}$$
$$\approx -\sigma^{-2} \ ,$$

where to leading order in $s$, $\sigma^2$ is again:

$$\sigma^2 \equiv N + s^2 N^2 \ , \tag{A14}$$

i.e., the variance in equation (A4). Also, to leading order,

$$\frac{d^3}{dn^3} Log[P[n;p,r]] = -(n+r-1)^{-2} + n^{-2}$$
$$\approx (1-p^2) N^{-2} \quad (n = n_0) \tag{A15}$$
$$= (1 + 2Ns^2)/\sigma^4.$$

Finally, expanding as a Taylor's series we have:

$$Log[P[n;p,r]] \approx constant - \tfrac{1}{2\sigma^2}(n-n_0)^2 + \tfrac{1}{6}(1+2Ns^2) \sigma^{-4}(n-n_0)^3. \tag{A16}$$

$$P[n;p,r] \approx \frac{1}{\sqrt{2\pi}\sigma} Exp[-\tfrac{1}{2\sigma^2}(n-n_0)^2] \ (1 + \tfrac{1}{6}(1+2Ns^2) \sigma^{-4}(n-n_0)^3) \ , \tag{A17}$$

where exponentiation of the cubic term has been expanded as perturbation. $P[n;p,r]$ has been normalized to unity:

$$\int_{-\infty}^{+\infty} P[n;p,r]\, dn = 1, \tag{A18}$$

within the accuracy of equation (A6) at $n = n_0$, noting that the cubic term in equation (A17) does not contribute.

Comparison can now be made between both the negative-binomial distribution and its approximate equation (A17) as well as the real data reported by Ogden, 1982, by computing cumulative distributions of the function pivot[$n,N$] (equation (2)) as in Fig. 1 which summarizes the experimental data.

Plotted in Fig. A1 is the cumulative distribution of pivot[$n,N$] at $s = 20\%$. The discrete points refer to exact computation according to equation (A1) for $N$ equal to 5, 15, and 25. The approximation (equation (A17)) is seen in Fig. A1 as virtually a single curve for $N = 5, 15,$ and $25$.

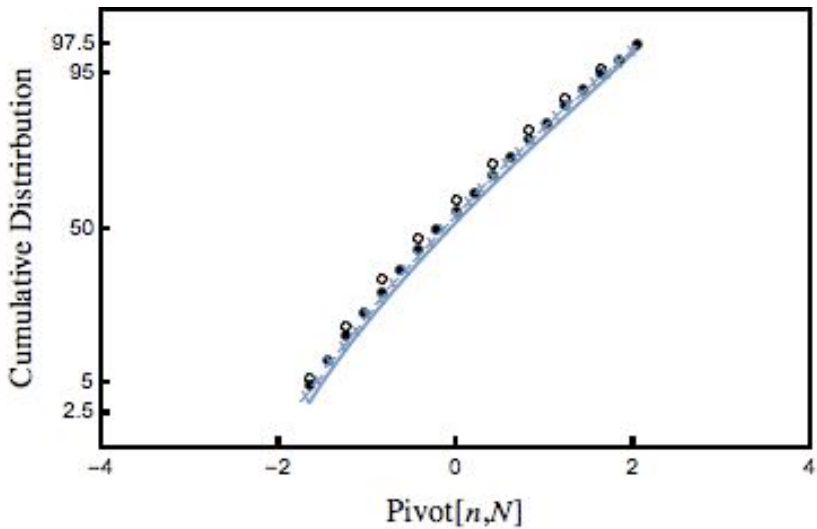

Fig. A1. Cumulative distribution of pivot[$n,N$] at $s = 20\%$. Open circles refer to $N = 5$, closed to $N = 15$, and "x" to $N = 25$, all computed directly according to the negative-binomial distribution (equation (A1)). The solid curve was computed from the approximation (A17).

The closeness of the results from the approximate equation (A17) and the exact negative binomial distribution (1) is apparent as is also the near $N$-independence in the distribution of pivot[$n,N$] as assumed and found by Ogden, 1982. Also, the results are consistent with the real data of Figure 1, considering that the large tail in the experimental data is subject to large uncertainty due to the small number of data points in this region.

*Quantile estimates*

We are now in position for computing quantiles. The quantile at level β (e.g., 5%) is found by solving for $\Delta$ in:

$$\beta = \sum_{n=0}^{n_0+z_\beta\sigma+\Delta} P[n;p,r] \tag{A19}$$

$$\approx \{\int_{-\infty}^{z_\beta} \tfrac{1}{\sqrt{2\pi}} e^{-\tfrac{1}{2}z'^2} dz' + \tfrac{1}{2}\tfrac{1}{\sqrt{2\pi}} e^{-\tfrac{1}{2}z_\beta^2}\} + \Delta/\sigma \times \tfrac{1}{\sqrt{2\pi}} e^{-\tfrac{1}{2}z_\beta^2} + \tfrac{1}{6}(1+2Ns^2)/\sigma \int_{-\infty}^{z_\beta} \tfrac{1}{\sqrt{2\pi}} z'^3 e^{-\tfrac{1}{2}z'^2} dz',$$

where $z_\beta$ is the normal-distribution quantile (e.g., $z_{0.05} = -1.645$ and $z_{0.95} = +1.645$). The term in brackets is $\sum_{n=0}^{n_0+z_\beta\sigma} \frac{1}{\sqrt{2\pi}\sigma} Exp[-\tfrac{1}{2}z^2]$, approximated as an integral by adding in ½ of the end point---the inverse of trapezoidal integration. The 2nd term is the first-order term in an expansion of $\int_{-\infty}^{z_\beta+\Delta/\sigma} \tfrac{1}{\sqrt{2\pi}} e^{-\tfrac{1}{2}z'^2} dz'$ in $\Delta/\sigma$. The 3rd term is

$$\int_{-\infty}^{z_\beta+\Delta/\sigma} \tfrac{1}{\sqrt{2\pi}} e^{-\tfrac{1}{2}z'^2} z^3 \tfrac{1}{6}(1+2Ns^2)/\sigma \, dz' \approx \int_{-\infty}^{z_\beta} \tfrac{1}{\sqrt{2\pi}} e^{-\tfrac{1}{2}z'^2} z^3 \tfrac{1}{6}(1+2Ns^2)/\sigma \, dz'$$ to lowest (0th)

order in $\Delta/\sigma$. Now the integral inside the bracket equals $\beta$, which therefore cancels $\beta$ on the left-hand side. Secondly, the cubic integral of the 3rd term is explicitly:

$$\int_{-\infty}^{z_\beta} \tfrac{1}{\sqrt{2\pi}} z'^3 e^{-\tfrac{1}{2}z'^2} dz' = -\tfrac{1}{\sqrt{2\pi}} e^{-\tfrac{1}{2}z_\beta^2}(2+z_\beta^2),$$

regardless of the sign of $z_\beta$. Therefore, finally, the above equation can be solved for $\Delta$ as:

$$\Delta = -\tfrac{1}{2} + \tfrac{1}{6}(1+2Ns^2)(2+z_\beta^2). \tag{A20}$$

The quantile $n_\beta$ is then:

$$n_\beta = 1 + IntegerPart[n_0 + z_\beta\sigma + \Delta], \tag{A21}$$

$$n_0 = N(1-s^2) - \tfrac{1}{2} \quad \text{(see equation (A12) above)}$$

where the function *IntegerPart* (which truncates fractions) expresses the step-function character of the quantiles. Equation (A21) implies that continuous (averaging) quantiles may be defined as:

$$\begin{aligned}\bar{n}_\beta &= \tfrac{1}{2} + n_0 + z_\beta \sigma + \Delta \\ &= N + z_\beta \sigma + \tfrac{1}{6}(z_\beta^2 - 1)(1 + 2Ns^2),\end{aligned} \quad (A22)$$

as illustrated in Figures (2-4).

*Poisson distribution* (s = 0)

As is easily seen in Figure 5, $(\bar{n}_\beta - N)/\sigma$ is significantly non-constant for *s* < 20%. This reflects the fact that $(\bar{n}_\beta - N)/\sigma$ is not analytic at the point $s \to 0$ (Poisson) and $N^{-1/2} \to 0$. For dealing with the Poisson distribution, rather than using the pivot approximation, equation (12) can be solved directly for *N*, as setting *s* equal to zero results in a quadratic equation in $\sqrt{N}$. Then in this special case:

$$\sqrt{N[n_\beta, z_\beta]} = -\tfrac{1}{2} z_\beta + \tfrac{1}{2}\sqrt{\tfrac{1}{3}z_\beta^2 + \tfrac{2}{3} + 4n_\beta}. \quad (A23)$$

Confidence limits (neglecting the fluctuations (illustrated below) from the discreteness of *n*) on *N* are then given in terms of the function $N[n, z_\beta]$:

$$\Pr[N[n, z_\beta] < N < N[n, z_{1-\beta}]] = 2\beta - 1. \quad (A24)$$

Equation (A24) can be compared to the more accurate traditional approximate confidence limits given by a relation of Poisson and chi-square quantiles (Johnson et al., 1993):

$$\Pr[\chi^2_{2n\ 1-\beta} < N < \chi^2_{2(n+1)\ \beta}] \approx 2\beta - 1. \quad (A25)$$

In order to judge the accuracy of the above expressions when applied to discrete random variables, many simulations were carried out. At each of mean number *N* = 5.1, 5.1, ..., 30, a random number generator produced 10,000 Poisson-distributed values for checking the above inequalities. Results are given in Figures (A2, A3). The analogue to equation (18) is:

$$\lim[N] \approx \beta \pm \frac{1/2}{(\sqrt{N} + \tfrac{1}{3}z_\beta)\sqrt{2\pi}\ Exp[z_\beta^2/2]}. \quad (A26)$$

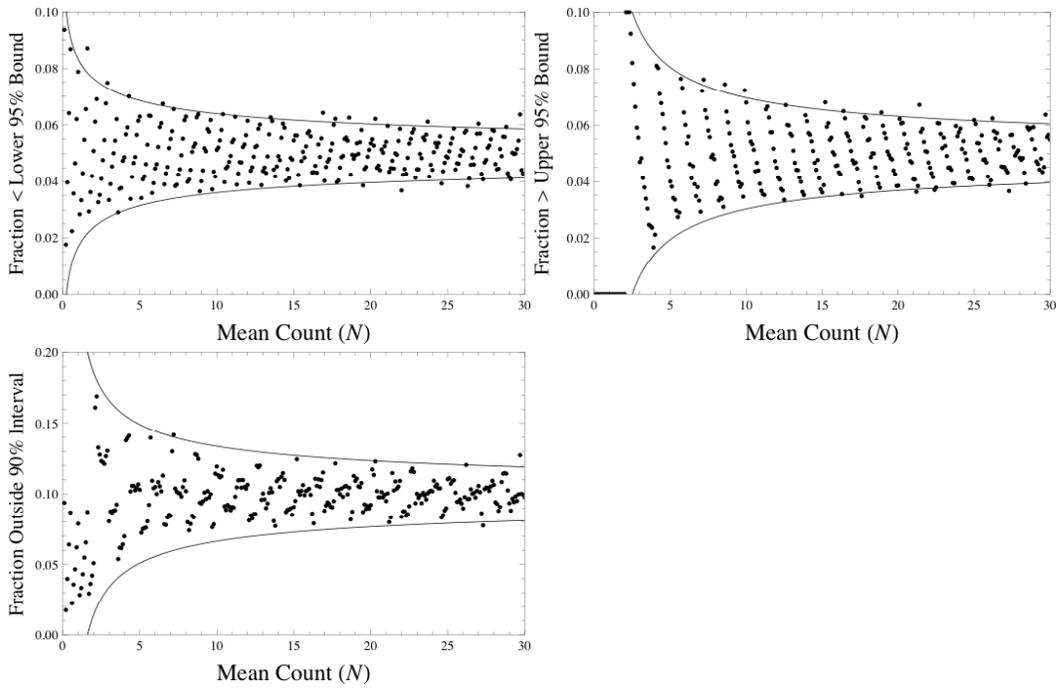

Fig. A2 Coverage probabilities with intervals computed following Stirling's approximation. Solid curves are approximate limits on the discrete variation (equation (A26)). Each point represents 10,000 simulated Poisson-distributed values.

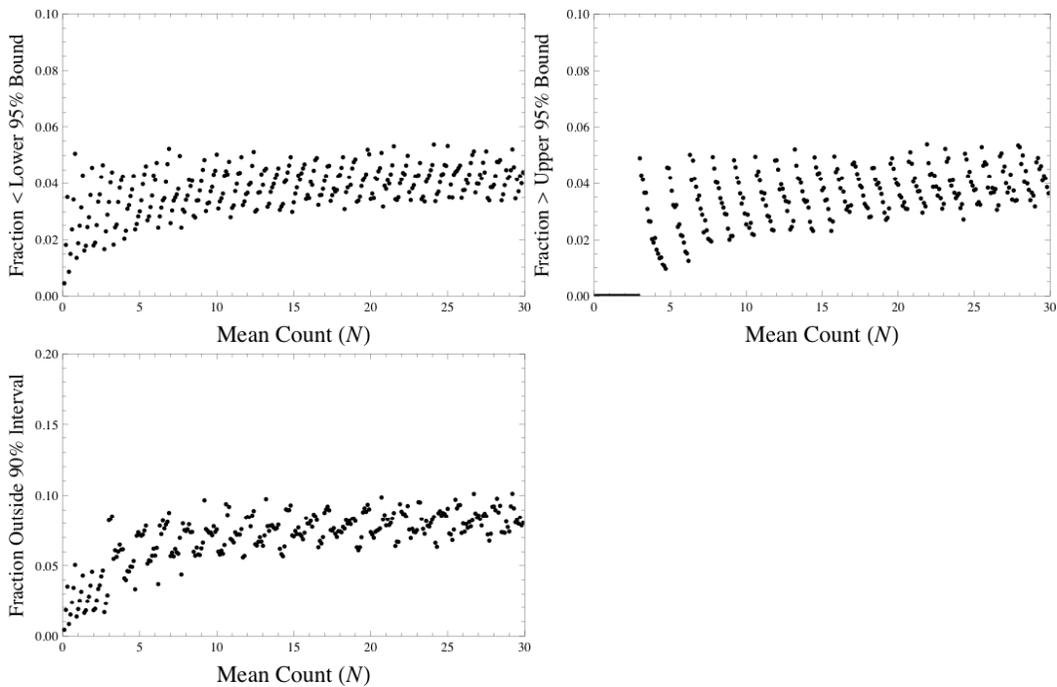

Fig. A3 Coverage probabilities with chi-square intervals. Each point represents 10,000 simulated Poisson-distributed values.

Note that the chi-square results are biased in a way that the probability of $N$ falling outside the intended (5% or 10%) confidence intervals is controlled conservatively. On the contrary, the Stirling approximation leads naturally to unbiased limits controlled in the mean. Of course, since the discrete scatter is strictly limited, conservative control could be set up if needed.